\documentclass[fleqn,10pt]{wlscirep}
\usepackage[utf8]{inputenc}
\usepackage[T1]{fontenc}
\usepackage{multirow}
\usepackage{xcolor,colortbl}
\usepackage{caption}
\usepackage{subcaption}

\usepackage{multicol}

\usepackage{adjustbox}

\usepackage{float}

\title{Estimation  of absorbed dose in organs  using Waza-AriV2 web-based software in head CT scans}

\author[1,4,*]{Juan Miguel Lozano}
\author[3,5]{H. Alejo-Martinez}
\author[2,4]{Edison Salazar}
\author[5]{M.E. Castellanos}

\affil[1]{Universidad del Valle, Physics department, Cali, Colombia}
\affil[2]{Universidad del Valle, Radiodiagnostics Department, Cali, Colombia}
\affil[3]{Hospital Universitario del Valle 'Evaristo García' E.S.E., Diagnostic Imaging Department, Cali, Colombia}
\affil[4]{Clinica Imbanaco Grupo Quirónsalud, Diagnostic Imaging Department, Cali, Colombia}
\affil[5]{Pontificia Universidad Javeriana, Medical Physics Program, Bogota, Colombia}

\affil[*]{Lozano.juan@correounivalle.edu.co}

\begin{abstract}
\textcolor{blue}{\textbf{ARISING FROM}} K. Yamashita et al. Nature. Scientific Reports, 11, 5435 (2021) \href{https://doi.org/10.1038/s41598-021-85060-5}{doi:10.1038/s41598-021-85060-5} (2021).

\end{abstract}
\begin{document}

\flushbottom
\maketitle
%
%
\thispagestyle{empty}

\begin{multicols}{2}

\hspace{-0.5cm}Recently, Yamashita et al\cite{yamashita}, reported direct measurements of ionizing radiation doses to individual organs in postmortem (using OSL dosimetry) that were exposed in the most frequent clinical CT scan protocols, corresponding to the whole body, head, thorax and abdomen. The authors\cite{yamashita} compare the experimental results with the values calculated in the Waza-AriV2 software\cite{waza1,waza2,waza3}, which is a system based on the monte carlo method that performs the estimation of organ doses from the exposure parameters used in the CT equipment. However, the results of the absorbed doses calculated in each examination had significant differences  with respect to the doses measured directly in postmortem. In this communication, we made a calculation under the correct interpretation in one of the acquisition parameters which leads to a significant discrepancy between the estimated and measured values reported by Yamashita et al. We also make the comparison between the Waza-AriV2\cite{waza1,waza2,waza3} and NCICT\cite{ncict} software.\\

\hspace{-0.5cm}The effective dose of radiation received by patients undergoing diagnostic CT examinations can be classified in the zone of low absorbed doses\cite{UN}. On the other hand, the effective dose is associated with the relative risk of radiation exposure for the entire population, however, caution should be exercised when estimating individual radiological risk from this magnitude. The absorbed dose in organs is more associated with the estimation of individual risk\cite{martin}. Direct dose measurements in individual organs, in six postmortem exposed in different clinical CT scan protocols, were performed by Yamashita et al\cite{yamashita}. The authors reported the exposure parameters of the CT scans carried out in a SOMATOM Emotion 16-slice (SIEMENS) and the average dose values in individual organs. These measured values were compared with those calculated with the software Waza-AriV2\cite{waza1,waza2,waza3}, which performs a dose estimation in organs based on the monte carlo method using computational anthropomorphic phantoms according to the parameters of the CT scan. The absorbed doses calculated with the software were significantly different from those reported experimentally, in particular, in the head examinations.
The absorbed dose measured in the brain for the head CT scan was 31.18 mGy, while the calculated one was 123.31 mGy, the absorbed dose measured in the lens was 29.95 mGy, being the estimated equal to 117.2 mGy, which corresponds to a  relative difference of 295.5$\%$ and 291.3$\%$, respectively. This has some important implications because according to reports from the United States, approximately half of the collective effective dose administered by imaging procedures is due to CT studies\cite{NCRP}. Some studies suggest that repeated CT exposures to head, neck, and sinus are related to an increased risk of cataracts\cite{yuan,zammit}. The ICRP has published threshold dose values for detectable lens opacities of 0.5 Sv for acute or prolonged exposures\cite{ICRP}. In this way, for patients who receive multiple CT studies, the accumulated lens doses could be of the order and even exceed the said  threshold. Other studies\cite{lechel} have shown that organ dose calculation software is increasingly used in dose tracking. In a comparative study of four available software, it was found that all were within 35$\%$ of variation with respect to the doses in organs that are inside the scan region\cite{demattia}. For this reason, the significant differences found in the experimental results published by K. Yamashita et al\cite{yamashita}, could lead to an overestimation or underestimation of the dose in organs, and, therefore, an increase in the uncertainty when estimating the risk of the radiological effects.
\\

\hspace{-0.5cm}The radiation dose in patients undergoing CT examinations is determined by parameters related to the equipment. Among them are beam filtration, beam collimation, detector configuration, image acquisition mode, kV, tube current, rotation time, slice thickness, pitch, and so forth. The charge on the x-ray tube in a CT is an exposure parameter that is defined as the product of the current in the x-ray tube (in units of milliamperes, mA) and the rotation time of the tube (in seconds, s). With the arrival of multislice scanners, some manufacturers (including Siemens) introduced an additional parameter called the effective mAs defined as:

\begin{equation}
\label{eqn:eq1}
{mAs}_{{eff}}=\frac{{mAs}}{{pitch}},
\end{equation}

\hspace{-0.5cm}where the pitch relates the advance of the table per rotation to the total width of the collimated beam.\\

\begin{figure}[!h]%
    \centering
    \subfloat[\centering Dose display shown at the end of a Siemens CT Head scan]{{\includegraphics[width=13cm]{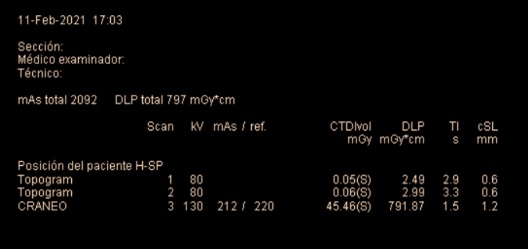} }}%
    \qquad
    \subfloat[\centering Reconstructed image of the same CT scan study ]{{\includegraphics[width=13cm]{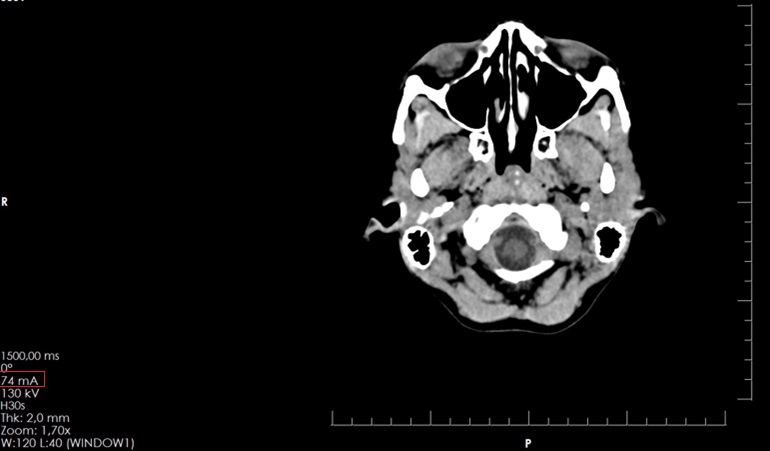} }}%
    \caption{\textbf{(a)} Dosimetric parameters displayed by the CT, the mAs shown is actually the effective mAs, {/}ref is the effective reference mAs from which the current will be modulated using the CareDose4D. \textbf{(b)} Reconstructed image of the same study \textbf{(a)} in which the value of the tube current in the z position corresponding to the acquisition shown can be observed in the lower left corner of the display (74 mA).}%
    \label{tab:fig1}%
\end{figure}

\hspace{-0.5cm}In the CT Siemens equipment, the mAs$_{\text{eff}}$ is displayed on the CT operator console during scan parameter setup and post-scan, as seen in Figure \ref{tab:fig1}a. Organ dose calculation software requires the choice of the computational phantom that best represents the patient and the type of CT, as well as the information of the scan parameters configured during the study. In the ‘table 1’ reported by Yamashita et al\cite{yamashita}, they present the irradiation conditions in the CT protocols, the tube current is mentioned in units of mAs, this unit does not correspond to the tube current (mA). In order to understand the parameters used and reproduce the absorbed doses in organs reported by the authors\cite{yamashita} using waza-ariV2\cite{waza1,waza2,waza3},  we first took a sample of 20 patients from our institutions in a weight range similar to those of the study in question (between 50 and 60 kg) that were exposed to head CT scans on a SOMATOM Scope 16-slice (SIEMENS) and a Somatom Go Now 16-slice (SIEMENS) under the same parameters reported by K. Yamashita et al\cite{yamashita}. Siemens Medical Solutions uses a current modulation system called CAREDose4D, which performs angular and longitudinal current modulation, adapted to different patient sizes and anatomical regions, for which the mAs$_{\text{eff}}$ 
of each study displayed on the console corresponds to an average mAs$_\text{eff}$ of all acquisitions (figure \ref{tab:fig1}a). The current modulation in the tube is carried out based on a reference value of mAs
 (mAs$_{\text{ref}}$) selected by the user according to the anatomy of the patient. In the sample of selected patients, the average mAs$_{\text{eff}}$ of all the studies was 215 mAs, which is close to the value reported by Yamashita et al\cite{yamashita} in the ‘table 1’ for head scans, which was 220 mAs. In this way, we intuit that the value entered in the software, for the calculation of organ doses, is the mAs$_\text{eff}$ instead of tube current (in units of mA).

\hspace{-0.5cm}If we carry out the respective correction, calculating the real value of the tube current from equation (\ref{eqn:eq1}) we have that

\begin{equation}
\label{eqn:eq2}
{mA}=\frac{{mAs}_{{eff}}\times {pitch}}{{t}_{r}},
\end{equation}

\hspace{-0.5cm}where t$ _\text{r} $ corresponds to the rotation time of the tube.\\
Yamashita et al\cite{yamashita}, mention that none of the reduction techniques was carried out, including current modulation (no CareDose4D), therefore the tube current must be constant. For the head protocol, the authors reported a rotation time of 1.5 s and a pitch of 0.55, for which the value corresponding to the tube current (80.67 mA) was obtained from equation (\ref{eqn:eq2}). This tube current is close to the current displayed on the equipment screen for a specific acquisition in one of our patients, when the study images are reconstructed (figure \ref{tab:fig1}b). This value is relevant to estimate the organ doses using the monte carlo software Waza-AriV2\cite{waza1,waza2,waza3} 
and is shown in 'table 2' by K. Yamashita et al\cite{yamashita}, for the head scan protocol. Using the same parameters of the said table, in which the values that we supply to the software Waza-AriV2\cite{waza1,waza2,waza3} can be seen in table \ref{tab:table2}, We replicate the results shown in the ‘table 3’ of the K. Yamashita et al\cite{yamashita} publication; these results can be seen in table \ref{tab:table3}.
\end{multicols}

\begin{table}[h!]
\centering
\begin{tabular}{ccccc}
\hline
\multicolumn{5}{|c|}{Head scan}                                                                         \\ \hline
Tube voltage (kV) & Effective mAs/ Ref & Pitch factor & Beam collimation (mm) & Rotation time (seconds) \\
130               & 212 / 220            & 0.55         & 16 × 0.6              & \multicolumn{1}{c}{1.5}
\end{tabular}
\caption{\label{tab:table1}Default parameters in a CT scan Siemens Somatom Scope 16-silce with a scan length of 174mm in default head clinical protocol.}
\end{table}

\newpage

\newpage


\begin{table}[H]
\centering
 \begin{adjustbox}{max width=\textwidth}
\begin{tabular}{|c|c|c|}
\hline
\multicolumn{3}{|c|}{Head scan Waza-AriV2.}                                                       \\ \hline
Manufacturer/Scaner Model   & Siemens                            & \multicolumn{1}{|c|}{Emotion 16} \\
\hline
Filter                      & \multicolumn{2}{|c|}{Head}                                            \\
Tube Potential              & \multicolumn{2}{|c|}{130 kV}                                          \\
Rotation Time               & \multicolumn{2}{|c|}{1.5 s}                                           \\
Pitch Factor                & \multicolumn{2}{|c|}{0.55}                                            \\
Beam width                  & \multicolumn{2}{|c|}{16x0.6mm}                                        \\
Gender                      & \multicolumn{2}{|c|}{Male}                                            \\
Phantom                     & \multicolumn{2}{|c|}{Adult Optional Phantom}                          \\
\hline
Scan Type                   & \multicolumn{1}{|c|}{Head}           & Routine Head (helical)         \\
\multirow{2}{*}{Scan Range} & \multicolumn{1}{|c|}{Begin Position} & 1835 mm                        \\
                            & \multicolumn{1}{|c|}{End Position}   & 1675 mm                        \\
                            \hline
AEC                         & \multicolumn{2}{|c|}{Off}                                             \\
\textbf{Tube Current}       & \multicolumn{2}{|c|}{\textbf{220 mA}}                                 \\
Optional Phantom            & \multicolumn{2}{|c|}{On}                                              \\
Height                      & \multicolumn{2}{|c|}{158.9}                                           \\
Weight                      & \multicolumn{2}{|c|}{51.6}                                            \\
CTDI Phantom Size           & \multicolumn{2}{|c|}{16cm}                                            \\
SSDE                        & \multicolumn{2}{|c|}{Off} \\
\hline
\end{tabular}
\end{adjustbox}
\caption{\label{tab:table2}Parameters given to the Waza-AriV2 web-based software to replicate the results obtained by K. Yamashita et al.}

\begin{tabular}{cccc}

Organs                                    & \begin{tabular}[c]{@{}c@{}}Waza-AriV2 \\ replicated organ dose\\  (mGy)\end{tabular} & \begin{tabular}[c]{@{}c@{}}Reported   Waza-AriV2\\  simulated Organ dose\\  by K.Yamashita et al. (mGy)\end{tabular} & \begin{tabular}[c]{@{}c@{}}Difference \\ between the two organ \\ doses (mGy)\end{tabular} \\
\cellcolor[HTML]{D7E4BC}Gonad             & \cellcolor[HTML]{FFFFFF}0                                                           & \textless{}0.01                                                                                               & \textless{}0.01                                                                            \\
\cellcolor[HTML]{D7E4BC}Prostate / uterus & \cellcolor[HTML]{FFFFFF}0                                                           &                                                                                                               &                                                                                            \\
\cellcolor[HTML]{D7E4BC}Urinary bladder   & \cellcolor[HTML]{FFFFFF}0                                                           &                                                                                                               &                                                                                            \\
\cellcolor[HTML]{D7E4BC}Colon             & \cellcolor[HTML]{FFFFFF}0.01                                                        & \textless{}0.01                                                                                               & \textless{}0.01                                                                            \\
\cellcolor[HTML]{D7E4BC}Small intestine   & \cellcolor[HTML]{FFFFFF}0                                                           & \textless{}0.01                                                                                               & \textless{}0.01                                                                            \\
\cellcolor[HTML]{D7E4BC}Kidney            & \cellcolor[HTML]{FFFFFF}0.02                                                        & 0.01                                                                                                          & 0.01                                                                                       \\
\cellcolor[HTML]{D7E4BC}Pancreas          & \cellcolor[HTML]{FFFFFF}0.02                                                        &                                                                                                               &                                                                                            \\
\cellcolor[HTML]{D7E4BC}Gall bladder      & \cellcolor[HTML]{FFFFFF}0.02                                                        &                                                                                                               &                                                                                            \\
\cellcolor[HTML]{D7E4BC}Stomach           & \cellcolor[HTML]{FFFFFF}0.03                                                        &                                                                                                               &                                                                                            \\
\cellcolor[HTML]{D7E4BC}Spleen            & \cellcolor[HTML]{FFFFFF}0.04                                                        &                                                                                                               &                                                                                            \\
\cellcolor[HTML]{D7E4BC}Adrenals          & \cellcolor[HTML]{FFFFFF}0.03                                                        &                                                                                                               &                                                                                            \\
\cellcolor[HTML]{D7E4BC}Liver             & \cellcolor[HTML]{FFFFFF}0.05                                                        & 0.03                                                                                                          & 0.02                                                                                       \\
\cellcolor[HTML]{D7E4BC}Heart             & \cellcolor[HTML]{FFFFFF}0.19                                                        &                                                                                                               &                                                                                            \\
\cellcolor[HTML]{D7E4BC}Lungs             & \cellcolor[HTML]{FFFFFF}0.27                                                        & 0.16                                                                                                          & 0.11                                                                                       \\
\cellcolor[HTML]{D7E4BC}Breast            & \cellcolor[HTML]{FFFFFF}0.2                                                         &                                                                                                               &                                                                                            \\
\cellcolor[HTML]{D7E4BC}Esophagus         & \cellcolor[HTML]{FFFFFF}0.36                                                        &                                                                                                               &                                                                                            \\
\cellcolor[HTML]{D7E4BC}Thymus            & \cellcolor[HTML]{FFFFFF}0.3                                                         &                                                                                                               &                                                                                            \\
\cellcolor[HTML]{D7E4BC}Thyroid           & \cellcolor[HTML]{FFFFFF}1.57                                                        & 0.8                                                                                                           & 0.77                                                                                       \\
\cellcolor[HTML]{D7E4BC}Salivary glands   & \cellcolor[HTML]{FFFFFF}33.67                                                       &                                                                                                               &                                                                                            \\
\cellcolor[HTML]{D7E4BC}Oral cavity       & \cellcolor[HTML]{FFFFFF}15.56                                                       &                                                                                                               &                                                                                            \\
\cellcolor[HTML]{D7E4BC}Out of Thorax     & \cellcolor[HTML]{FFFFFF}101.58                                                      &                                                                                                               &                                                                                            \\
\cellcolor[HTML]{D7E4BC}Lens              & \cellcolor[HTML]{FFFFFF}122.75                                                      & 117.2                                                                                                         & 5.55                                                                                       \\
\cellcolor[HTML]{D7E4BC}Brain             & \cellcolor[HTML]{FFFFFF}132.57                                                      & 123.31                                                                                                        & 9.26                                                                                       \\
\cellcolor[HTML]{D7E4BC}Lymphaden         & \cellcolor[HTML]{FFFFFF}5.77                                                        &                                                                                                               &                                                                                            \\
\cellcolor[HTML]{D7E4BC}Muscle            & \cellcolor[HTML]{FFFFFF}3.41                                                        &                                                                                                               &                                                                                            \\
\cellcolor[HTML]{D7E4BC}Skin              & \cellcolor[HTML]{FFFFFF}10                                                          &                                                                                                               &                                                                                            \\
\cellcolor[HTML]{D7E4BC}Bone              & \cellcolor[HTML]{FFFFFF}56.65                                                       &                                                                                                               &                                                                                            \\
\cellcolor[HTML]{D7E4BC}Active marrow     & \cellcolor[HTML]{FFFFFF}8.39                                                        &                                                                                                               &                                                                                            \\
                                                                                                                                                                                                                                                                                                                                            
\end{tabular}

\caption{\label{tab:table3}Comparison between the simulated results obtained and the published results by K. Yamashita et al. Using the Waza-AriV2 software.}

\end{table}

\begin{table}[H]
\begin{tabular}{cccc}
Organs                                    & \begin{tabular}[c]{@{}c@{}}Waza-AriV2 \\ corrected organ dose\\  (mGy)\end{tabular} & \begin{tabular}[c]{@{}c@{}}Postmortem \\  reported organ dose \\  by K.Yamashita et al. (mGy)\end{tabular} & \begin{tabular}[c]{@{}c@{}}Difference \\ between the two organ \\ doses (mGy)\end{tabular} \\
\cellcolor[HTML]{D7E4BC}Gonad             & \cellcolor[HTML]{FFFFFF}0               & 0.01 $\pm$ 0.0004                                 & \multicolumn{1}{c}{0.01 $\pm$ 0.0004}                                    \\
\cellcolor[HTML]{D7E4BC}Prostate / uterus & \cellcolor[HTML]{FFFFFF}0               &                                                                &                                                                          \\
\cellcolor[HTML]{D7E4BC}Urinary bladder   & \cellcolor[HTML]{FFFFFF}0               &                                                                &                                                                          \\
\cellcolor[HTML]{D7E4BC}Colon             & \cellcolor[HTML]{FFFFFF}0               & 0.05 $\pm$ 0.03                                                & \multicolumn{1}{c}{0.05 $\pm$ 0.03}                                      \\
\cellcolor[HTML]{D7E4BC}Small intestine   & \cellcolor[HTML]{FFFFFF}0               & 0.05 $\pm$ 0.03                                                & \multicolumn{1}{c}{0.05 $\pm$ 0.03}                                      \\
\cellcolor[HTML]{D7E4BC}Kidney            & \cellcolor[HTML]{FFFFFF}0.01            & 0.04 $\pm$ 0.03                                                & \multicolumn{1}{c}{0.03 $\pm$ 0.03}                                      \\
\cellcolor[HTML]{D7E4BC}Pancreas          & \cellcolor[HTML]{FFFFFF}0.01            &                                                                &                                                                          \\
\cellcolor[HTML]{D7E4BC}Gall bladder      & \cellcolor[HTML]{FFFFFF}0.01            &                                                                &                                                                          \\
\cellcolor[HTML]{D7E4BC}Stomach           & \cellcolor[HTML]{FFFFFF}0.01            &                                                                &                                                                          \\
\cellcolor[HTML]{D7E4BC}Spleen            & \cellcolor[HTML]{FFFFFF}0.01            &                                                                &                                                                          \\
\cellcolor[HTML]{D7E4BC}Adrenals          & \cellcolor[HTML]{FFFFFF}0.01            &                                                                &                                                                          \\
\cellcolor[HTML]{D7E4BC}Liver             & \cellcolor[HTML]{FFFFFF}0.02            & 0.13 $\pm$ 0.09                                                & \multicolumn{1}{c}{0.11 $\pm$ 0.09}                                      \\
\cellcolor[HTML]{D7E4BC}Heart             & \cellcolor[HTML]{FFFFFF}0.07            &                                                                &                                                                          \\
\cellcolor[HTML]{D7E4BC}Lungs             & \cellcolor[HTML]{FFFFFF}0.1             & 0.76 $\pm$ 0.9                                                 & 0.66 $\pm$ 0.9                                                           \\
\cellcolor[HTML]{D7E4BC}Breast            & \cellcolor[HTML]{FFFFFF}0.07            &                                                                &                                                                          \\
\cellcolor[HTML]{D7E4BC}Esophagus         & \cellcolor[HTML]{FFFFFF}0.13            &                                                                &                                                                          \\
\cellcolor[HTML]{D7E4BC}Thymus            & \cellcolor[HTML]{FFFFFF}0.11            &                                                                &                                                                          \\
\cellcolor[HTML]{D7E4BC}Thyroid           & \cellcolor[HTML]{FFFFFF}0.57            & 1.32 $\pm$ 0.70                                                & 0.75 $\pm$ 0.70                                                          \\
\cellcolor[HTML]{D7E4BC}Salivary glands   & \cellcolor[HTML]{FFFFFF}12.33           &                                                                &                                                                          \\
\cellcolor[HTML]{D7E4BC}Oral cavity       & \cellcolor[HTML]{FFFFFF}5.7             &                                                                &                                                                          \\
\cellcolor[HTML]{D7E4BC}Out of Thorax     & \cellcolor[HTML]{FFFFFF}37.22           &                                                                &                                                                          \\
\cellcolor[HTML]{D7E4BC}Lens              & \cellcolor[HTML]{FFFFFF}44.97           & 29.95 $\pm$ 3.84                                               & 15.02 $\pm$ 3.84                                                         \\
\cellcolor[HTML]{D7E4BC}Brain             & \cellcolor[HTML]{FFFFFF}48.57           & 31.18 $\pm$ 2.18                                               &     17.39 $\pm$ 2.18 
                                                                     \\
\cellcolor[HTML]{D7E4BC}Lymphaden         & \cellcolor[HTML]{FFFFFF}2.12            &                                                                &                                                                          \\
\cellcolor[HTML]{D7E4BC}Muscle            & \cellcolor[HTML]{FFFFFF}1.25            &                                                                &                                                                          \\
\cellcolor[HTML]{D7E4BC}Skin              & \cellcolor[HTML]{FFFFFF}3.66            & $\dagger$                & {\color[HTML]{242729}  $\dagger$} \\
\cellcolor[HTML]{D7E4BC}Bone              & \cellcolor[HTML]{FFFFFF}20.75           &                                                                &                                                                          \\
\cellcolor[HTML]{D7E4BC}Active marrow     & \cellcolor[HTML]{FFFFFF}3.08            &                                                                &                                                                          \\

\end{tabular}
\caption{\label{tab:table5}Comparison between the simulated results obtained with the corrected mA, using the Waza-AriV2 software and the published results by K. Yamashita et al. Measured in postmortem. $\dagger$ We did not consider the skin dose, because the postmortem measurements\cite{yamashita} were made only in the nipple and Umbilicus (outside the head scan region).}
\end{table}

 \begin{multicols}{2}

\begin{figure}[!htb]
\centering
\includegraphics[width=\linewidth]{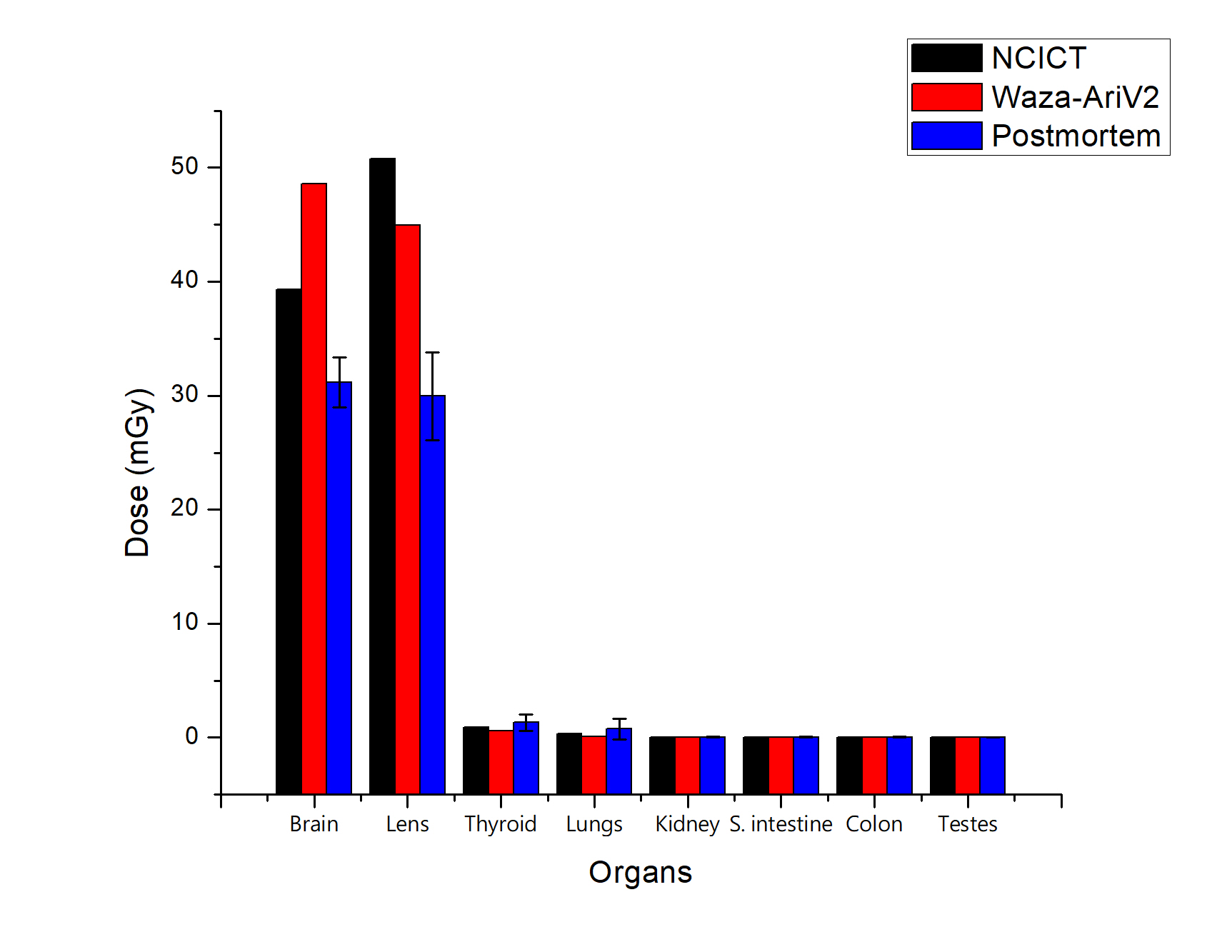}
\caption{Comparison of dose estimation in organs in two software (NCICT \& Waza-AriV2) and organ dose measured in postmortem reported by Yamashita et al.}
\label{fig:sof}

\end{figure}

\hspace{-0.5cm}It was observed (table \ref{tab:table3}) that the replicated values of absorbed dose in organs obtained by the Waza-AriV2\cite{waza1,waza2,waza3} compared with the published results, differ in the main irradiated organs in  9.26 mGy (brain) and 5.55 mGy (lens), 
as well as differences less than 0.01 mGy in organs outside the scan region. Those variations compared with the values reported by K Yamashita et al\cite{yamashita}, suggest that the tube current given in the software\cite{waza1,waza2,waza3} in order to estimate the organ doses corresponds to the effective mAs, value that is mentioned in 'table 1' reported by K. Yamashita et al\cite{yamashita}. The small variations may be due to the fact that the exact parameters supplied to the software by the authors are unknown, a male virtual phantom was also chosen in contrast to the female one, chosen by the authors\cite{yamashita}, under the argument that 2/3 of the total postmortem samples were males. On the other hand, the correction of the value of the tube current was done (80.67mA) from which we proceeded to estimate the organ doses, using the software\cite{waza1,waza2,waza3}, these doses were compared with the  values measured in postmortem reported by the authors K. Yamashita et al\cite{yamashita}, these findings were reported in table \ref{tab:table5}. The dose variations with the corrected tube current (mA) are of the same order as those measured by K Yamashita et al\cite{yamashita} in postmortem  (giving a better approximation) with differences in the main irradiated organs of 15.02 $\pm$ 3.84 mGy (lens) and 17.39 $\pm$ 2.18 mGy (brain). Organ differences outside the scan region are less than 0.1 mGy. The differences between the measured values and those calculated from the software\cite{waza1,waza2,waza3} can be attributed to systematic errors induced by taking the average height and weight of various postmortems, without discrimination by gender, and to variations with respect to computational phantoms that have standard characteristics. This generates changes in the absorbed doses in organs calculated by the Waza-AriV2 software\cite{waza1,waza2,waza3}. Additionally, the NCICT software\cite{ncict} was used to compare calculated doses (figure \ref{fig:sof}). The other CT scan protocols evaluated should also be reviewed following a similar methodology.
\newpage

\hspace{-0.5cm}The preliminary results obtained show that the absorbed dose values calculated with the two software (figure \ref{fig:sof}) seem to overestimate the dose with respect to the postmortem measurements in the head scan, however, they represent a good approximation in contrast to the results of Yamashita et al.\cite{yamashita}. More detailed studies on individual postmortem dosimetry and the comparison of these doses with the different software should be carried out.
 
\end{multicols}

 \
 
 \begin{multicols}{2}

\section*{Methods}
We observed the effective mAs in the dose display of two CT's (Somatom Scope 16-slice, Somatom Go Now 16-silce (SIEMENS)) with the same parameters in twenty patients between 50 and 60 kg of our institutions to compared with the tube current reported  by K. Yamashita et al\cite{yamashita}. We also used the web-based software Waza-AriV2\cite{waza1,waza2,waza3} to replicate the data reported by K. Yamashita et al.\cite{yamashita} and the equation (\ref{eqn:eq2}) from which we correct the value of the tube current. In addition we used the NCICT software\cite{ncict} to compared the reviewed results with the Waza-ariV2\cite{waza1,waza2,waza3} and the postmortem organ doses reported by K. Yamashita et al.\cite{yamashita}

\section*{Data availability}

For our analyses, we used the ‘Data of the table 1, Figure 4, and table 3 of K. Yamashita et al.\cite{yamashita}’ to replicate the data using the Waza-AriV2 web based software\cite{waza1,waza2,waza3} and the data of the table\ref{tab:table2}. We also used the ‘Data of the table 2, of K. Yamashita et al.\cite{yamashita}’ to compared the corrected tube current organ doses estimated with Waza-AriV2\cite{waza1,waza2,waza3} and the NCICT software\cite{ncict}. Any other reasonable data requests should be addressed to J.M.L.

\bibliography{sample}

\section*{Author contributions statement}

J.M.L., H.A.M and E.S. contributed equally to this
work. M.E.C provided Comments and approved the final work.

\section*{Competing interests}

The authors declare no competing interests.

 \end{multicols}

\end{document}